\documentclass[epj]{svjour}
\usepackage{graphics}
\newcommand{\be}{\begin{equation}}
\newcommand{\ee}{\end{equation}}
\newcommand{\bea}{\begin{eqnarray}}
\newcommand{\eea}{\end{eqnarray}}
\newcommand{\bean}{\begin{eqnarray*}}
\newcommand{\eean}{\end{eqnarray*}}

\newcommand{\gapproxeq}{\lower
.7ex\hbox{$\;\stackrel{\textstyle >}{\sim}\;$}}
\newcommand{\lapproxeq}{\lower
.7ex\hbox{$\;\stackrel{\textstyle <}{\sim}\;$}}

\newcommand{\nrightarrow}{\mbox{$ \rightarrow\hspace{-0.4cm}\backslash\hspace{0.4cm}$}}
\def\3bar{$\bar {\hbox{\bf 3}}$}

\begin{document}
\title{Selection rules and quark correlations in the $N^*$ resonance
spectrum }
\author{Qiang Zhao\inst{1,2} \and Frank E. Close\inst{3}% etc
}                     % Do not remove
\institute{Institute of High Energy Physics, Chinese Academy of
Sciences, Beijing 100049, P.R. China \and Department of Physics,
University of Surrey, Guildford, GU2 7XH, United Kingdom \and
Rudolf Peierls Centre for Theoretical Physics, University of
Oxford, Keble Rd., Oxford, OX1 3NP, United Kingdom}
\date{Received: date / Revised version: date}
\abstract{ A ``$\Lambda$ selection rule" for $N^*$ resonances in
the presence of QCD mixing effects is identified. Due to the QCD
mixing, excitations of {\bf 20}-plets are possible in SU(6). We
show that this selection rule is useful for classifying PDG states
at $N=2$, and for clarifying whether strongly correlated diquarks
survive for $L > 0$.
 % end of PACS codes
} %end of abstract
\maketitle
\section{Quark model and selection rules}
\label{intro}

Although it has been about forty years since the quark model was
first applied to the problem of baryon resonances, it is still not
well established whether three constituent quarks or a
quark-diquark effective degrees of freedom are needed in the
description of the baryon spectrum. In the recent years,
significant progresses on the photo-nuclear reactions as a probe
for the internal structure of nucleon and nucleon resonance have
been made in experiment, which provide not only stringent
constraints on theoretical phenomenologies but also novel insights
into the strong QCD dynamics in this challenging regime.

A standard and phenomenologically successful assumption common to
a large number of papers in the quark model is that photon
transitions are additive in the constituent
quarks~\cite{close-book,cko,fkr}. This assumption also underlies
models of hadronic production and decay in the sense that when
$q_1q_2q_3 \to  [q_1q_2q_i] + [q_3 \bar{q_i}]$, the quark pair
$q_1q_2$ are effectively spectators and only $q_3$ is involved in
driving the transition. Such approximations lead to well known
selection rules, which have proved useful in classifying
resonances\cite{close-book}. We adopt this approximation as a
first step and show that within it there is a further selection
rule that appears to have been overlooked in the literature. We
shall refer to this as the ``$\Lambda$ selection rule" and show
how it may help classify $N^*$ resonances~\cite{zhao-close}.

The standard SU(6)$\otimes$O(3) wavefunction can be constructed
from three fundamental representations of group $S_3$:
\begin{eqnarray}
SU(6): \ \ & {\bf 6}\otimes{\bf 6}\otimes{\bf 6} &= {\bf 56}_s
+{\bf 70}_\rho + {\bf 70}_\lambda + {\bf 20}_a ,
\end{eqnarray}
where the subscripts denote the corresponding $S_3$ basis for each
representation, and the bold numbers denote the dimension of the
corresponding representation. The spin-flavor wavefunctions can be
expressed as $|{\bf N}_6, ^{2S+1} {\bf N}_3\rangle $, where ${\bf
N}_6$ (={\bf 56, \ 70} or {\bf 20}) and ${\bf N}_3$ (={\bf 8, \
10}, or {\bf 1}) denote the SU(6) and SU(3) representation and $S$
stands for the total spin. The SU(6)$\otimes$O(3) (symmetric)
wavefunction is
\begin{equation}
|\mbox{SU(6)}\otimes\mbox{O(3)}\rangle = |{\bf N}_6, \ ^{2S+1}{\bf
N}_3, \ N, L, J\rangle \ ,
\end{equation}
where explicit expressions follow the convention of Isgur and
Karl~\cite{isgur-karl,ik-78,ik70mix}.

The basic rules follow from application of the Pauli exclusion
principle to baryon wavefunctions together with an empirically
well tested assumption that electroweak and strong decays are
dominated by single quark transitions where the remaining two
quarks, or diquark, are passive
spectators~\cite{capstick-roberts}. As a consequence, it leads to
a correlated vanishing transition matrix element between $N^*$ of
[{\bf 70}, \ $^4 8$] and [{\bf 56}, \ $^2 8$] in $N^* \to \Lambda
K$ or $\Lambda K^*$. This follows because the $[ud]$ in the
$\Lambda$ has $S=0$ and in the spectator approximation,  the
strangeness emissions in $N^* \to \Lambda K$ or $\Lambda K^*$,
 the spectator $[ud]$ in the $N^*$ must also be in
$S_{[ud]}=0$, whereby such transitions for the $N^*$ of $[{\bf
70}, \ ^4 8]$ with $S_{[ud]}=1$ are forbidden.

This ``$\Lambda$ selection rule", which appears to have been
overlooked in the literature, seems to be useful for resolving the
underlying transition dynamics and probing the structure of
excited $N^*$. We note the well-known Moorhouse selection
rule~\cite{moorhouse} which states that transition amplitudes for
$\gamma p$ to all resonances of representation $[{\bf 70}, \ ^4
8]$, such as $D_{15}(1675)$, must be zero due to the vanishing
transition matrix element for the charge operator. In contrast,
the $\Lambda$ selection rule applies to both proton and neutron
resonances of $[{\bf 70}, ^4 8]$. We also note that
$\Lambda^*[{\bf 70}, \ ^4 8] \nrightarrow \bar{K}N$ has been
discussed in Ref.~\cite{hey,fai77}. But the source and generality
of the $\Lambda$ section rule does not seem to have been
noted~\cite{fai77}.

\section{Recognition of the $\Lambda$ selection rule}

An immediate application of the selection rules is to the
$D_{15}(1675)$, which is in $[{\bf 70}, ^4 8]$. According to the
Moorhouse selection rule, the amplitudes for $\gamma p \to D_{15}$
should vanish. However, the experimental values are not zero,
though they are small.  Non-zero amplitudes arise from QCD mixings
induced by single gluon exchange in the physical
nucleon~\cite{ikk}. The effective interaction
\bea
H_{FB} &= & \frac{2\alpha_s}{3m_i m_j} \left[\frac{8\pi}{3}{\bf
S}_i\cdot{\bf S}_j\delta^3({\bf r}_{ij}) \right. \nonumber\\ & &
\left.+\frac{1}{r_{ij}^3}\left(\frac{3({\bf S}_i\cdot {\bf
r}_{ij})({\bf S}_j\cdot {\bf r}_{ij})}{r_{ij}^2}  - {\bf
S}_i\cdot{\bf S}_j \right) \right] \label{hfb}
\eea
induces significant mixings between the $^2${\bf 8} and $^4 ${\bf
8} in the {\bf 56} and {\bf 70}~\cite{ik70mix} and the nucleon
wavefunction becomes~\cite{ikk}
\begin{equation}
|N\rangle = 0.90|^2S_S\rangle -0.34|^2S_{S'}\rangle -0.27|^2S_M
\rangle -0.06|^4 D_M\rangle \ , \label{nmix}
\end{equation}
where subscripts, $S$ and $M$, refer to the spatial symmetry in
the $S$ and $D$-wave states for the nucleon internal wavefunction.
Thus, the $O(\alpha_s)$ admixtures at $N=2$ comprise a 34\% in
amplitude excited {\bf 56} and 27\% {\bf 70} each with $L=0$ and
6\% {\bf 70} with $L=2$.  The following points can be learned by
applying the rules to compare with the experimental observations:

i) Due to the QCD mixing in the wavefunction of the nucleon, the
Moorhouse selection rule is violated. The {\bf 70} admixture
quantitatively agrees with the most recent data~\cite{pdg2006} for
the $\gamma p \to D_{15}$ amplitudes, neutron charge radius and
$D_{05} \to \bar{K} N$~\cite{ikk}. The results assume that mixing
effects in the $D_{15}$ are negligible relative to those for the
nucleon\cite{ikk}: this is because there is no $[{\bf 70}, ^2 8;
L^P = 1^-]$ state available for mixing with the $D_{15}$, and the
nearest $J=5/2$ state with negative parity is over 500 MeV more
massive at $N=3$.  The leading $O(\alpha_s)$ amplitude for $\gamma
p \to D_{15}$ is dominantly driven by the small components in the
nucleon and the large component in the $D_{15}$~\cite{ik-78} for
which $\Delta N =1$.

ii) The $\Lambda$ selection rule remains robust, or at least as
good as the Moorhouse rule even at $O(\alpha_s)$. This is because
in the context of the diquark model, admixtures of $[ud]$ with
spin one, which would violate the selection rule, are only
expected at most to be 20\% in amplitude~\cite{ik-80}, to be
compared with 27\% for the nucleon in Eq.~(\ref{nmix}). Thus
decays such as $D_{15} \to K\Lambda$ will effectively still vanish
relative to $K\Sigma$; for the $D_{15}(1675)$ the phase space
inhibits a clean test but the ratio of branching ratios for the
analogous state at $N=2$, namely $F_{17}(1990) \to K\Lambda : K
\Sigma$, may provide a measure of its validity.

iii) For $\gamma n\to D_{15}$, where the Moorhouse selection rule
does not apply, the amplitudes are significantly large and
consistent with experiment~\cite{pdg2006}. However, due to the
$\Lambda$ selection rule, the $D_{15}^0 \nrightarrow K^0\Lambda$
which makes the search for the $D_{15}$ signals in $\gamma N\to
K\Lambda$ interesting. An upper limit of $B.R.< 1\%$ is set by the
PDG~\cite{pdg2006} which in part may be due to the limited phase
space; a measure of the ratio of branching ratios for $K \Lambda :
K \Sigma$ would be useful. The $F_{17}(1990)$, which is the only
$F_{17}$ with $N=2$, is an ideal candidate for such a test, which
may be used in disentangling the assignments of the positive
parity $N^*$ at the $N=2$ level.

\section{Excitation of 20-plets}

The QCD admixture of $[{\bf 70}, ^2 8, 2, 0, 1/2]$ in the nucleon
wavefunction enables the excitation of {\bf 20}-plets. There has
been considerable discussion as to whether the attractive forces
of QCD can cluster $[ud]$ in color {\bf $\bar{3}$} so tightly as
to make an effective bosonic ``diquark" with mass comparable to
that of an isolated quark. Comparison of masses of $N^*(u[ud])$
and mesons $u\bar{d}$ with $L \geq 1$ support this hypothesis of a
tight correlation, at least for excited states~\cite{wilc04}. If
the quark-diquark dynamics is absolute, then SU(6)$\otimes$O(3)
multiplets such as $[{\bf 20},1^+]$ cannot occur. The spatial
wavefunction for {\bf 20} involves both $\rho$ and $\lambda$
degrees of freedom; but for an unexcited diquark, the $\rho$
oscillator is frozen. Therefore, experimental evidence for the
excitations of the {\bf 20} plets can distinguish between these
prescriptions.

In Ref.~\cite{zhao-close} the transition amplitudes for the lowest
$[{\bf 20},1^+]$ to {\bf 70}-plets are given. It is also shown
that the amplitudes are compatible with the Moorhouse-
selection-rule-violating amplitudes in $\gamma p\to D_{15}$. Thus,
a 27\% {\bf 70} admixture in the nucleon has potential
implications for resonance excitation that may be used to look for
{\bf 20}-plets. Nevertheless, additional $P_{11}$ and $P_{13}$
from representation {\bf 20} automatically raise questions about
the quark model assignments of the observed $P_{11}$ and $P_{13}$
states, among which $P_{11}(1440)$, $P_{11}(1710)$, and
$P_{13}(1720)$ are well established resonances, while signals for
$P_{13}(1900)$ and $P_{11}(2100)$ are quite poor~\cite{pdg2006}.

\section{Implications from the present experimental data}

At $N=2$ in the quark model a quark-diquark spectrum allows $[{\bf
56}, 0^+]$, $[{\bf 56}, 2^+]$, and $[{\bf 70}, 0^+]$$[{\bf 70},
2^+]$. If all $qqq$ degrees of freedom can be excited,
correlations corresponding to $[{\bf 20}, 1^+]$ are also possible.
As shown in Ref.~\cite{zhao-close}, without {\bf 20}-plets, most
of the observed states can fit in the SU(6) scheme. However, there
are still a lot of problems and controversies. For instance,
neither of the $P_{13}(1710 \\ /1900)$ fit easily with being pure
{\bf 56} or {\bf 70} states; states of $[{\bf 70}, ^2 8, 2,2,J]$
are still missing, and another $P_{13}$ and $F_{15}$ are needed.

When the QCD mixing effects are included the agreement improves,
in that small couplings of $^4N$ states to $\gamma p$ are
predicted, in accord with data. However, the implication is the
added complexity that an additional $P_{11}$ and $P_{13}$
correlation in [{\bf 20},$1^+$] is allowed. Most immediately this
prevents associating the $P_{11}(1710)$ as [{\bf 70,$0^+$}] simply
on the grounds of elimination of alternative possibilities. Thus
we now consider what are the theoretical signals and what does
experiment currently say.

\begin{table}
\caption{Helicity amplitudes for the $P_{11}(1710)$ and
$P_{13}(1720)$ with all the possible quark model assignments for
them. The data are from PDG~\cite{pdg2006}, and numbers have a
unit of $10^3\times$GeV$^{1/2}$. }
\label{tab-2}       % Give a unique label
\begin{tabular}{c|c|c|c}
\hline\noalign{\smallskip}
&  $P_{11}(1710)$&  $P_{13}(1720)$ &  $P_{13}(1900)$  \\
& $A^p_{1/2}$ & $A^p_{1/2}$ \ \ \ $A^p_{3/2}$ & $A^p_{1/2}$  \ \ \
$A^p_{3/2}$\\
 \noalign{\smallskip}\hline\noalign{\smallskip}
$[{\bf 56}, ^2 8; 2^+]$ & * & 100 \ \ \ \ \ 30 & 110 \ \ \ \ \ 39  \\
$[{\bf 70}, ^2 8; 0^+]$ & 32 & * \ \ \ \ \ \ \ * & * \ \ \ \ \ \ \ *  \\
$[{\bf 70}, ^2 8; 2^+]$ & * & $-71$ \ \ \ \ \ $-21$ & $-78$ \ \ \ \ \ $ -28$ \\
$[{\bf 70}, ^4 8; 0^+]$ & * & 17 \ \ \ \ \ \ \ 29 & 14 \ \ \ \ \ \ \ 24 \\
$[{\bf 70}, ^4 8; 2^+]$ & $-8$ & 7 \ \ \ \ \ \ \ 12 & 9 \ \ \ \ \ \ \ 16 \\
$[{\bf 20}, ^2 8; 1^+]$ & $-15$ & $-11$ \ \ \ \ \ $-18$ & $-10$ \ \ \ \ $-17$ \\
Exp. data & $9\pm 22$ & $18\pm 30$ \ $-19\pm 20$ & $-17$ \ \ \
$+31$ \\
 \noalign{\smallskip}\hline
\end{tabular}
% Or use
%\vspace*{5cm}  % with the correct table height
\end{table}

Qualitatively one anticipates $P_{13}(^4N)$ having a small but
non-zero coupling to $\gamma p$, the $\gamma n$ being larger while
the $K \Lambda$ decay is still forbidden. For the {\bf 20} states
$P_{11,13}(^2N)$ both $\gamma p$ and $\gamma n$ amplitudes will be
small and of similar magnitude. However, mixing with their
counterparts in {\bf 56} and {\bf 70} may be expected. In
Table~\ref{tab-2}, we list the helicity amplitudes for the
$P_{11}(1710)$, $P_{13}(1720)$ and $P_{13}(1900)$ with all the
possible quark model assignments and the mixing angles from
Eq.~(\ref{nmix}). The amplitudes for the $P_{11}$ and $P_{13}$ of
$[{\bf 20}, ^2 8, 2,1, J]$ are the same order of magnitude as the
Moorhouse-violating $\gamma p\to D_{15}(1675)$~\cite{zhao-close}.
For the $P_{11}(1710)$ all three possible configurations  have
amplitudes compatible with experimental data. For $P_{13}(1720)$,
assignment in either $[{\bf 56}, ^2 8; 2^+]$ or $[{\bf 70}, ^2 8;
2^+]$ significantly overestimates the
data~\cite{close-li,li-close} for $A^p_{1/2}$ if it is a pure
state.

Table~\ref{tab-2} shows that the presence of {\bf 20} cannot be
ignored, should be included in searches for so-called ``missing
resonances", and that a possible mixture of the {\bf 20}-plets may
lead to significant corrections to the results based on the
conventional {\bf 56} and {\bf 70}. This raises a challenge for
experiment, whether one can eliminate the {\it extreme}
possibility that $P_{11}(1710)$ and $P_{13}(1720)$ are consistent
with being in ${\bf 20}$ configurations. There are already
qualitative indications that they are not simply {\bf 56} or {\bf
70}. Their hadronic decays differ noticeably from their sibling
$P_{11}(1440)$: compared with the $P_{11}(1440)$ in {\bf 56} for
which $\Gamma_T \sim 350 $ MeV with a strong coupling to $N\pi$,
their total widths are $\sim 100$ MeV, with $N\pi$ forming only a
small part of this.

Some further implications due to the {\bf 20}-plets excitations
can be learned here:

i) For a state of {\bf 20}-plet, as ${\bf 20} \nrightarrow {\bf
56} \otimes {\bf 35}$, whereas ${\bf 20} \rightarrow {\bf 70}
\otimes {\bf 35}$ is allowed, decay to $N \pi$ will be allowed
only through the ${\bf 70}$ admixtures in the nucleon. This makes
the $P_{11}(1710)$ and $P_{13}(1720)$ extremely interesting as not
only are their total widths significantly less than the
$P_{11}(1440)$ but the dominant modes for $P_{11}(1710)$ and
$P_{13}(1720)$ are to $N\pi\pi$, which allows a possible cascade
decay of  ${\bf 20}\to N^*({\bf 70})\pi\to N\pi\pi$.

ii) The $\Lambda$ selection rule is useful for classifying the
$P_{11}$ and $P_{13}$ in either $[{\bf 56}, ^2 8]$ and $[{\bf 70},
^2 8]$, or $[{\bf 70}, ^4 8]$ and $[{\bf 20}, ^2 8]$, by looking
at their decays into $K \Lambda$ and/or $K^*\Lambda$. The $[{\bf
70}, ^4 8]$ decays to $K\Lambda$ will be suppressed relative to $K
\Sigma$ for both charged and neutral $N^*$.

iii) The Moorhouse selection rule can distinguish $[{\bf 70}, ^4
8]$ and $[{\bf 20}, ^2 8]$ since the $[{\bf 70}, ^4 8]$ will be
suppressed in $\gamma p$ but sizeable in $\gamma n$, while the
$[{\bf 20}, ^2 8]$ will be suppressed in both.

iv) $J/\psi \to \bar{p} + N^*$ is a further probe of $N^*$
assignments, which accesses {\bf 56} in leading order and {\bf 70}
via mixing while {\bf 20} is forbidden. Hence for example $J/\psi
\to \bar{p} + (P_{11}:P_{13})$ probes the {\bf 56} and {\bf 70}
content of these states. Combined with our selection rule this
identifies $J/\psi \to \bar{p} + K \Lambda$ as a channel that
selects the {\bf 56} content of the $P_{11}$ and $P_{13}$.

In summary, with interest in $N^*$ with masses above 2 GeV coming
into focus at Jefferson Laboratory and accessible at BEPC with
high statistic $J/\psi\to \bar{p}+ N^*$, the $\Lambda$ selection
rule should be useful for classifying baryon resonances and
interpreting $\gamma N \to K\Lambda$, $K^*\Lambda$, $K\Sigma$ and
$K^*\Sigma$. A coherent study of these channels may provide
evidence on the dynamics of diquark correlations and the presence
of {\bf 20}-plets, which have hitherto been largely ignored.


\end{document}